# Archives, archival bond, and digital representation: A case study with the International Image Interoperability Framework

**Martin Critelli**

martin.critelli@univ-amu.fr

Aix-Marseille University

**Additional Declarations:** No competing interests reported.

## Abstract

Within the archival sector, digitization has long been a strategic initiative to ensure greater availability of historical documents. In recent years, the promotion of guidelines and standards, combined with technological advancements, has established methodologies and best practices and developed tools to facilitate massive digitization projects. However, despite the availability of technological solutions and guidelines, digitization is intended mostly to scan documents and make the outcome images available online. This practice can be problematic in representing the complex fonds structure made of relations, the archival bond that establishes the natural ordering of documents into archival units. This is particularly relevant when the fonds also has a multimedia component, such as an audiovisual component, that is often reproduced on different platforms disconnected from textual documents. This article addresses the challenges linked to digitization in the archival sector and proposes a methodological framework for representing fonds with respect to their native organization. For this purpose, the International Image Interoperability Framework (IIIF) is employed to configure a specific model that respects the archive's hierarchical structure. In particular, this model is configured to maintain the archival bond and enhance the resource's semantic aspect to make the IIIF model semantically interoperable. To demonstrate the adaptability of the framework to the archival domain, in this work, the "PCI-Unitelefilm" fonds of the Fondazione Archivio Audiovisivo del Movimento Operaio e Democratico (AAMOD) served as the case study.

## Introduction

Traditionally, in the archival sector, digitization has always played a strategic role in ensuring access to documents and sharing the memory contained within them (Şentürk 2014; Azim et al. 2018; Zherdeva et al. 2020), "actively promoting and guaranteeing the development of culture" (Istituto Centrale per la Digitalizzazione del Patrimonio Culturale – Digital Library 2022). The availability of international standards (2010) and guidelines (Aly and Andrew 2023) significantly helps define criteria, benchmark models, and procedures for successfully digitizing archives. Furthermore, advances in the Information and Communications Technologies (ICT) sector and the development of advanced text recognition techniques such as optical character recognition (OCR) and handwritten text recognition (HTR) have undoubtedly contributed to large-scale digitization efforts, making documents machine-readable and more accessible (Terras 2022; Spina 2023; Nockels 2025; Romein et al. 2025). However, given the complexity of archive structures, simply applying these technologies is not enough to ensure full representation within a digital environment. An uncontrolled approach to digitization aimed only at creating online access can lead to a proliferation of images of documents without proper context. In this scenario, the archival fonds does not reflect the network of relations, expressed by the archival bond, which made its creation and subsequent sedimentation in the historical archive possible. If this occurs, archives lose their identity and essence, transforming into mere repositories (Castellucci 2017). This issue becomes even more apparent when digitized archives lack proper descriptions that highlight these relations among documents, files, series, or other fonds. For this reason, it is necessary to ask: what are the criteria that ensure that the provenance principle is respected when digitizing archival fonds?

Moreover, if, in the analog context, archivists are in charge of ensuring that the provenance principle is respected (Lodolini 1995), should it not be the same in the digital environment? These are necessary questions to be asked to ensure that digital is not just a container into which scanned copies of documents can be poured, but instead allows the archive to be represented *tout court*.

The situation worsens further if we consider archives with multimedia documents, such as audiovisual documents, which are often reproduced on different digital platforms. Thus, it is essential to understand that the knowledge, history, and culture preserved in archives are expressed through various document types, and their digitization raises questions about how to make them accessible, investigable, and usable without losing context. Beyond technical questions concerning funding availability, the tools to employ, and the resolution of digitized objects, it is also necessary to focus on the methodological aspects of managing the archive's digital representation. Over the past 15 years, the development and continuous updates of the International Image Interoperability Framework (IIIF) have contributed to the online sharing of archival and librarian material. Compared with classic digital libraries, IIIF was created with the original purpose of offering a creative and innovative environment; today, it constitutes one of the main solutions chosen by institutions and private companies for the digital representation of cultural material. Despite IIIF's potential for interoperability and advanced visualization, in the archival sector, it is often applied without exploiting its functionalities. This is even more remarkable than the librarian sector, characterized by a greater number of projects in which IIIF is applied to emphasize domain features, such as the Digital Vatican Library (Manoni 2023) or the Alphabetica project (Ministero della Cultura 2021). The risk of this trend is that it may not fully utilize the framework's capabilities to convey the complexity and information richness of archival material, specifically in its hierarchical organization and depth of description. As a consequence, IIIF is reduced to a static digital library in which images of digitized documents that are deprived of context are uploaded. However, the progressive updates from the IIIF community reveal the possibility of adapting the framework to the archival structure.

Thus, the present article is based on the following research questions (RQs) with the aim of overcoming the problem of losing the archive's identity in digital representation.

**RQ1**. Is IIIF sufficient to represent the complex archival structure and its relation network?

**RQ2**. How can the semantic level within the IIIF model be increased to guarantee the interoperability of archival descriptions?

**RQ3**. Can a single platform be used to present diverse multimedia content and enable users to analyze and investigate it?

[1] Commonly pronounced “Triple-Eye-Eff”.

# Methods

To address the above RQs, this work uses the Fondazione Archivio Audiovisivo del Movimento Operaio e Democratico (AAMOD) archive as a case study. In particular, this article presents a IIIF model configuration specifically designed for the representation of archival fonds that include both textual and audiovisual material. From a visual perspective, this work aims to demonstrate the potential of IIIF to highlight the hierarchical structure of archival fonds and the relationships among documents, files, and series. With respect to semantic interoperability, this work focuses on integrating standardized archival descriptions into the IIIF model, enabling their sharing and integration with other information systems. By incorporating texts, images, and audiovisuals into a single model, this work aims to provide a methodology for leveraging existing and open-source resources, such as IIIF, to preserve the archive's network of relationships.

The article first provides an overview of the IIIF framework, focusing on methods of resource description and international application cases in the archival context[2]. The second part moves into the details of the case study, presenting the characteristics of the IIIF models specific to the archival fonds structure. The description of the model is accompanied by screenshots to show a few samples of how it is visually accessible. The discussion section comments on the outcome of the work according to the RQs, and finally, this work ends with some conclusions focused on the proposed model and further applications.

# Description of archival resources within the IIIF standard

The IIIF standard was developed in 2011 to create a digital environment where cultural resources, especially manuscripts, can be shared at high resolution without significant economic and IT costs (Salarelli 2017). Unlike traditional, static digital libraries, this standard enables the development of an "interoperable" environment where resources are linked with their metadata and users can interact with them, utilizing various functionalities instead of just viewing digitized images (Snydman et al. 2015). The IIIF object includes a syntactic structure defined by the Presentation API (International Image Interoperability Framework), generating a JSON-LD object that can be easily and quickly shared via the HTTP(S) protocol and displayed in specific viewer applications. This structure is organized hierarchically into elements that can be arranged to represent either a single item or a collection. The main elements of this structure are as follows:

- **Collection**. This is the apex item of the IIIF structure and is configured to be an ordered list of Manifests or other Collections.
- **Manifest**. A Collection's sub-item that carries all the information related to the represented object.
- **Canvas**. A Manifest's sub-item is defined as the virtual container of the object.

As the framework's name suggests, interoperability is one of the main characteristics that distinguish IIIF, making it a suitable solution for universities, libraries, archives, and other cultural institutions to share their resources. However, it has already been emphasized (De Vincentis and Critelli 2025) that this interoperability is explicated at the syntactic level, with less attention given to the semantic profile, leaving cultural curators free to describe digital resources. The choice to focus mainly on syntactic interoperability is to provide a unique environment where resources can be shared without constraints imposed by culturally tailored metadata application profiles. Therefore, cultural curators can describe their resources according to the IIIF metadata structure, a specific section organized into four categories distinguished as follows:

- Descriptive;
- Structural;
- Technical;
- Linking.

Within each of these classes, a set of mandatory and optional metadata elements is defined. With respect to the class of descriptive metadata, the <*metadata*> property can be used to include descriptive information. This element is designed to contain customizable label—value pairs following the JSON—LD standard. Entering information into the <*metadata*> property allows users to obtain syntactically interoperable data that, however, do not convey semantic meaning. As a result, these label-value pairs lack semantic significance. The lack of a semantic descriptive system in IIIF grants users flexibility in resource descriptions but also hinders standardization. This approach risks producing only human-readable data, which fail to meet the definition of metadata as "machine-understandable information about web resources or other items" (W3C), with no access to its semantic content. To address this gap, IIIF's current version includes the <*seeAlso*> property, enabling users to embed external resources such as ontologies and descriptive standards (International Image Interoperability Framework). Currently, this element allows partial refinement of a resource's semantics within IIIF. For archival resources, XML descriptions such as EAD (Encoded Archival Description) or RiC-O (Records in Context Ontology) are preferably noted.

From a visual standpoint, a IIIF object can be displayed on the web via specific viewers. Today, many software applications have been created, each differing in features, functionalities, or operating system compatibility[3]. Often, choosing a particular viewer or implementing it with additional plugins can be crucial for enhancing the user's access experience. Despite the widespread adoption of the framework and the development of increasingly advanced features, basic IIIF applications remain common in the archival sector, mainly for displaying scanned document images. This reflects archival institutions' interest in the framework and its recent rise in popularity within this domain. Compared with the library sector, where the framework has been quickly adopted for sharing digitized materials, we have only recently begun to see its broader adoption among archival institutions.

To better understand the current application status of the framework in the archival sector, this section examines some of the most well-known cases, highlighting the characteristics that distinguish each case.

In the Italian context, several significant projects have been realized, notably the Teca Digitale of the Archivio Centrale dello Stato (ACS) and the Portale Antenati by the Direzione Generale Archivi (DGA). These projects are certainly among the most ambitious digitization projects in Italy, given the vast number of archival fonds preserved and their national relevance. The ACS Teca Digitale project began in July 2022 and currently contains tens of thousands of digitized documents and various fonds made available to users via IIIF (Archivio Centrale dello Stato). The project has been realized through the support of the Media Library Online (MLOL) and is characterized by the "Stories" module (Media Library Online). This functionality offers users a personal space to create and organize a collection of digitized material. The default viewer offered is a customized basic version of Mirador 3 (Matienzo et al.), through which users can search for text within images (Fig. 1).

In the case of the DGA Portale Antenati project (Direzione Generale Archivi), a specific Mirador plugin is used to allow users to add "annotations" to digitized documents. This plugin was developed within the Mirador community to increase users' interaction with the item. These annotations are made directly on the item and can then be stored locally on the user's device or remotely via a server application such as the Simple Annotations Server (Robinson et al. 2020). Owing to the ability to make their own comments and annotations, this functionality may be particularly effective for scholars who need to analyze and investigate documents stored elsewhere.

In addition to the cases mentioned earlier, other private and institutional archives in Italy have adopted IIIF to share their documentary collections. One example is the Archivio Storico of the University of Bologna (Archivio Storico dell'Università di Bologna), where the digitization effort is ongoing. This archive contains files related to literary figures such as Roberto Roveresi and Pier Paolo Pasolini. It uses a basic version of Mirador 2 as the default viewer, without any extra features. Like the Teca Digitale project, the Archivio Storico della Gazzetta di Parma has also used MLOL to provide access to its historical archive via IIIF, utilizing the so-called "Storie" feature (Archivio Storico della Gazzetta di Parma). This project was designed to highlight the role of historical newspapers within the local community. Users navigate the Gazzetta di Parma Archive Portal through a straightforward, three-part story path that highlights key moments in the newspaper's history.

Moving to a broader perspective, in the international context, the British Library serves as a key reference for the application of IIIF to archival and library materials. The various partnerships established with other institutions and the launch of specific programs (British Library 2025) have enabled the British Library to expand its network of relationships with documentary material, facilitating research, access, and analysis activities. Users can therefore access resources through the prestigious library's general catalog, where the availability of IIIF implementations is indicated by the framework's specific icon (Fig. 2). Compared with the Italian cases described above, the British Library has chosen the Universal Viewer (UniversalViewer) as the default viewer for its collection. Along with the basic functionalities of deep zoom and rotation, resource metadata are viewable in the column next to the image (Fig. 3).

The IIIF has also been employed within the Biblissima portal to make cultural material from different institutions, such as the Bibliothèque Nationale de France, the Biblioteca Nacional Digital de Portugal, and the Bayerische Staatsbibliothek available online. Here, the portal stands out for its ambition to create a prototype capable of organizing and indexing individuals Manifests through their contained metadata, normalizing language strings and authority files via specific controlled lists such as VIAF, Wikidata, LoC, etc., and experimenting with new metadata extraction techniques for the automatic creation of Manifests starting from specific descriptive records (Biblissima). A peculiarity of the project is the use of the Change Discovery API (International Image Interoperability Framework) to increase the searchability and accessibility of individual digital objects across participating providers. Within the Biblissima database, users can search for content filtered by institutions or consult the entire general catalog. Moreover, users can create a personal collection by selecting their favorite objects (Fig. 4), which can then be viewed via the default tool Mirador 2.

The projects illustrated here demonstrate that IIIF has been progressively used to make historical archives and documents available online. The framework represents a good solution, in some instances also promoted by national guidelines, such as the Ministero della Cultura and Istituto per la digitalizzazione del patrimonio culturale - Digital Library (n.d.), to facilitate the management and sharing of digitized resources. Archivists may leverage the framework's features to enhance and widely disseminate the material they are tasked with preserving. Nevertheless, more can still be done to fully utilize the framework’s features to improve the quality of services for disseminating archival material. For example, suffice to say that there is an almost total absence of a proper descriptive apparatus that semantically supports the resource conveyed through the framework. Institutions rarely integrate the description of archival material in the IIIF model and, most importantly, return the hierarchical order of the fond.

The following section presents a case study illustrating the configuration of a IIIF model tailored explicitly for the archival sector. The main characteristics of this model enable users to explore archives in the digital environment, as they would in an analog environment.

**Case study: the application of the IIIF standard for Fondazione Archivio Audiovisivo del Movimento Operaio e Democratico archival fond.**

Building on the case studies discussed in the previous paragraph, the second part of this work illustrates a IIIF model capable of digitally representing archival fonds while respecting its hierarchical order in series, subseries, and files and thus the relations between documents. In addition, the work focuses on enhancing the semantic interoperability of the entire model to ensure that it can be used on the web and across information systems to retrieve information and its digital representation. For this work, the archive of the Fondazione Archivio Audiovisivo del Movimento Operaio e Democratico, therefore Fondazione, was selected as a test sample to configure a suitable IIIF model. The selection of Fondazione’s archive was based on three main criteria:

1. An archive composed of text and multimedia documents.

2. Online availability and open-access material.
3. Presence of archival material partially or fully described.

Fondazione, established in the late 1970s, still operates currently in the field of audiovisuals (cinema, TV, multimedia, movies, and photographic archives) for the promotion, collection, and dissemination of collective history and memories of Italian social movements and their key figures. To date, the Fondazione holds a considerable number of archival fonds, including the audiovisual fonds of the Confederazione Generale Italiana del Lavoro (CGIL) and the Partito Comunista Italiano (PCI)[4]. Fondazione has distinguished its archival heritage into four primary categories—film libraries, texts, audio libraries, and photographs—and counts over a thousand hours of recordings, hundreds of texts, and half a million images. For this work, among the archival fonds, it was decided to select a few samples from different categories within a single fonds to demonstrate IIIF's adaptability to archives.

The selected material is part of the "PCI-Unitelefilm" fond, which comprises documents from various series that focus on Italian emigration in the 1960s. The main details of the chosen fonds are as follows:

**Holding Institution:** Fondazione Archivio Audiovisivo del Movimento Operaio e Democratico

**Fonds:** PCI – Unitelefilm

**Date:** 1926 – 1980

**Creator:** Unitelefilm

Materials: Video, audio, pictures, and textual documents.

**Reference:** http://patrimonio.aamod.it/aamod-web/film/detail/IL8000000006/22/fondso-unitelefilm.html

In the following subsections, the methodological and procedural aspects of the work are explained, with a focus on the tools employed and the characteristics of the IIIF model. In particular, the first part highlights the semantic aspects of the work, describing the activities aimed at increasing the semanticity of the IIIF model. The second part focuses on the technical aspects of the IIIF model, emphasizing its structural composition for representing archival fonds.

## Enhancing the semanticity within IIIF

The intrinsic structure of IIIF is not intended to ensure semantic interoperability of the information provided by the user. In this work, the choice is to follow a dual path that channels information through the IIIF structure while also incorporating semantically valid information. To meet this objective, the <*metadata*> property was used to detail the foundation's information about the material. With respect to the semantic aspect, the metadata <*seeAlso*> were used to embed the corresponding finding aid in XML by normalizing the information according to the EAD standard. The use of a specific archival-domain standard, formatted in XML, was necessary to ensure full interoperability of the information at both the semantic and syntactic levels.

The primary use of the EAD finding aid is to standardize the information provided by the Fondazione. However, the information presented on the website focuses primarily on the context of the resource rather than the content. Possible consequences of this situation include the risk of failing to exploit the information potential and thus losing knowledge. To overcome this limitation, Deep Learning, Computer Vision, and Natural Language Processing (NLP) techniques were applied to automatically analyze document content and extract information, providing greater descriptive granularity. By applying these techniques, the goal is to leverage hidden knowledge across various document types. To achieve this purpose, object detection models were employed to analyze images and audiovisual files, recognizing everyday items that can serve as additional details in describing the files. For this activity, a pretrained YOLO v8x model (Redmon et al. 2016) was selected; it was trained on the Google Open Images Dataset (Google), which contains hundreds of labels. On the other hand, in textual files, NLP and topic-extraction techniques were used to extract entities related to places, locations, and subjects. This has been accomplished by using spaCy (Ines Montani et al. 2023), NLTK (Bird et al. 2009), and BERTopic (Grootendorst 2022).

The result of these processes is a set of terms related to different categories (i.e., items, topics, places, etc.) that can be used within the finding aid to increase the archival description of the fonds and their subclasses. For this purpose, since the EAD3 schema allows referencing thesauri and controlled lists in some elements, the extracted elements were normalized via two specific thesauri: the Nuovo Soggettario of the Biblioteca Nazionale Centrale di Firenze (BNCF) and the Virtual International Authority File (VIAF). The decision to use the Nuovo Soggettario of BNCF is motivated by the context of fonds; since it is Italian, this thesaurus would be better suited for standardizing terms. In combination with the Nuovo Soggettario, the VIAF was utilized to normalize authority records and geographical places because of its extensive international coverage. The normalized terms were then expressed via the <*controlAccess*> section in the EAD file (Fig. 5).

In this sense, the application of the EAD schema is crucial for guaranteeing the interoperability of archival information. Thus, IIIF is not only a “container” for representing digital objects but also a medium for sharing standardized information and for aggregating metadata (Freire et al. 2017).

## The configuration of the IIIF model for the representation of archives

The process of configuring the discussed IIIF model begins with selecting which version of the Presentation API to use. Since this model has to address both images and audiovisuals, Presentation API v3, the latest version, was employed because it is the only one that supports both document types.

Following the selection of the API to use, the second aspect concerns the modalities for combining the IIIF elements — Manifest, Collection, and Canvas — to represent the hierarchical structure of archives properly. The basic concept of this methodological procedure is to associate these elements with the different archival units and combine them to reflect the order and relationships that characterize an archive. To be clearer, starting from the bottom of the structure, the organization was accomplished as follows:

- **Manifest of the item**. This represents a single archival item, with its metadata aligned. In this case, the Canvas element is used to host the document's digital representation.
- **Manifest of the file**. This represents the archival file, which is aligned with its metadata. This element contains as many Canvases as documents to be represented and is used to organize and make the digital representations of the archival items immediately visible and accessible.
- **Collection of the file**. This represents the file aligned with its metadata. This element is used to group and order the Manifests of the archival items.
- **Collection of broader archival unit**. This represents the higher archival units of the series or subseries. This element is used to group and order narrower archival units.
- **Collection of the fonds**. This ultimate element represents the entire fonds. It orders and organizes the collection of series to allow exploration of the fonds in accordance with their physical organization.

What immediately emerges from this structure is that both Manifest and Collection were used to represent and describe archival files. This decision provides two different access options for the file units. The Manifest of the file focuses primarily on the file unit and its finding aid and provides an overview of the items contained here. Thus, by accessing this element, users can analyze the file and understand its content. Instead, the Collection of the file is also focused on archival items; indeed, it allows the investigation of their Manifests in alignment with the file's Manifest. In adhering to these criteria, the ultimate structure is configured to provide access to the different archival levels in alignment with their respective metadata. The schema in Fig. 6 represents the conceptual organization of the IIIF elements.

In line with the IIIF model's conceptual structure, as shown in Fig. 6 above, the operative phase is characterized by a bottom-up approach, beginning with the creation of Manifests for the archival items. This approach is preferable for configuring the model without risking the loss of information or disorganizing the higher elements. Therefore, starting from the Manifest of the item, it is then possible to configure the Manifest of the file and so on until the Collection of the fonds. Despite the tightness, this structure allows for complete and immediate access to both items and archival units, respecting the original archive’s organization and thus the archival bond.

The result of this process is therefore the JSON-LD file that can be displayed via the already mentioned IIIF-specific viewers. For this case study, a customized version of Mirador 3 with a video-annotation plugin (Geourjon) was chosen as the default viewer. From a user-experience perspective, the intuitive viewer interface simplifies model navigation, improving content usability. The image below (Fig. 7) shows the model's initial screen, which is configured according to the Mirador 3 settings.

The model’s starting screen opens at the highest level of fonds and displays a list of sublevels (series). From here, it is then possible to navigate the narrower elements of the fonds according to their structure. Moving into the details of the visual representation, the image below shows the viewer’s main screen options at the archival unit level, represented as a Manifest of the file (Fig. 8).

In this sample, the left sidebar is structured to report the following:

- name of the displayed item (“Current item”);
- current IIIF level (“Collection”);
- information provided in the IIIF <*metadata*> section (“Resource”);
- embedded information and Manifest URI (“Related Links”).

The information shown in the “Resource” section was gathered from Fondazione’s website. In contrast, the standardized information in the EAD-XML file is provided under “Related Links,” as highlighted in the red box in Figure 9. The latter section is structured to contain both the link to the resource's referring webpage and the Manifest URI. The availability of these items makes it easy for users to switch between the IIIF model and the Fondazione website quickly, and they can also use the URIs to display the model in other IIIF-compatible viewers. The center of the Mirador's screen is reserved for displaying the digital object. In the example mentioned, the object depicted is an archival item that constitutes the file. The bottom bar of the thumbnails provides access to the complete list of file items for users. Each of these thumbnails represents a Canvas of the displayed Manifest, intended to provide an overview of the file and simulate a "showcase" of its content.

From this level, it is then possible to move into the narrowest element, the Manifest of the item, by opening the list menu “Show Collection”, as highlighted by the blue box in Figure 9. From this menu, users can select the desired archival level to access the corresponding resource. The layout of the Manifest of the item is the same as the Manifest of the file, but in this case, the representation and metadata are totally inherent to the archival item. To increase user engagement, this version of Mirador 3 has been personalized with the video-annotation plugin. This option allows users to select, highlight, and make comments on specific temporal and spatial zones of audiovisual items.

[2] For the purposes of this article, only the necessary notions regarding the IIIF will be described given without repeating information widely discussed in literature. For a complete overview of the organization and structure of IIIF, we refer to: https://iiif.io/api/presentation/3.0/.

[3] For a full list of software, tools, and resources IIIF compliant, please visit: https://iiif.github.io/awesome-iiif/.

[4] For more information regarding the Fondazione’s history, please visit the page: https://aamod.it/

## Discussion

The proposed IIIF model configuration was finalized to represent the archival structure in the digital environment, maintaining the archival bond between documents. The model proposed here was configured using only open-source technologies, making it a feasible solution for most archival institutions with limited financial resources.

In relation to the proposed RQs, the following is possible:

**RQ1**. *Is IIIF sufficient to represent the complex archival structure and its relation network?*

The IIIF structure demonstrated great flexibility, particularly in the archival context. Indeed, the features of the main elements enabled a link between IIIF and archival elements. By exploiting these features, it was possible to associate archival units and items with the IIIF elements Collection, Manifest, and Canvas. This operation was necessary to attribute an archive in digital form that can be organized according to its physical sedimentation. Indeed, by following the syntactic schema of IIIF, it was then possible to reproduce the original fonds' organization and keep alive the archival bond on which it is based. In this structure, archival items can be represented in their context rather than as isolated objects.

**RQ2**. *How can the semantic level within the IIIF model be increased to guarantee the interoperability of archival descriptions?*

Notably, IIIF does not focus on semantics; instead, it allows users to be free to describe resources. In this case study, the <*seeAlso*> property was used to embed an external resource to increase the semantic interoperability of the resource reproduced via IIIF. Given the domain, applying the EAD standard assisted in this purpose by enabling information sharing across the network. Indeed, this is not a definitive solution, but given the current IIIF specifications, it might constitute a helpful methodology for sharing archives' digital representations aligned with standardized information.

**RQ3**. *Can a single platform be used to present diverse multimedia content and enable users to analyze and investigate it?*

The current IIIF Presentation API has clearly helped address this question by providing a structure capable of embedding both images and audiovisual content. In terms of visual aspects, the availability of advanced, creative tools makes IIIF a practical solution for sharing digitized materials. In this particular case, Mirador 3 and its plugins have enabled the configuration of an all-in-one tool that users can use to access and annotate the entire archival collection. Notably, Mirador allows users to view multiple objects simultaneously, allowing them to analyze records from the same or different files at once. This feature is useful because it allows users to review documents within the same environment without switching tabs (Fig. 9). While IIIF is sufficient to display the current documents preserved in historical archives, it is also important to consider that recent file types, such as 3D objects, technical images, medical images, and others, have emerged. To prepare for future needs, we must consider how to render these formats, taking into account key factors such as file size, computational resources, and the energy required to display them.

Finally, the last consideration concerns the use of computational methods to extract information and increase the descriptive granularity of documents. This was a secondary activity that was processed via pretrained models. In the case of object detection, the process has several limitations, which can be attributed to a lack of proper model fine-tuning to process such images and audiovisuals. For future applications, a suitable dataset can be used to refine the model and increase the accuracy of the results. With respect to text files, NLP proved practical for extracting certain entities, particularly dates and locations, to enhance the informativeness of documents in archival finding aids. However, this final process did not constitute the core of the work; instead, it served as a means to retrieve additional information. Further work is planned to focus on this aspect, making a model available that can investigate and extract information from this type of documentation.

## Conclusions

In the work presented here, IIIF was proposed as a solution for digitally representing an archival collection, preserving its physical structure and maintaining the network of relationships among the different archival items. This goal was achieved by showcasing a IIIF configuration tested on a part of the fonds “PCI-Unitelefilm” of Fondazione Archivio Audiovisivo del Movimento Operaio e Democratico as a case study. Using the core elements of the IIIF Presentation API, the model accurately depicted the hierarchical structure of the fonds. This model is specifically designed to keep text and audiovisual files together, preventing their dispersion across multiple platforms and thus preserving archival connections. Being an archivist requires respect for this fundamental principle to ensure that archives retain their historical and cultural significance over time. This responsibility does not end with digitization; it continues, especially as more archival institutions increasingly make their materials accessible online. In this context, scholars, researchers, and even general users would greatly benefit from having access to organized fonds online rather than collections of digitized images that lack connections to the original context. From an archivist's perspective, this solution eliminates the need to use multiple platforms to share different materials, creating a unified environment for accessing and examining documents. For example, an archival institution can adopt this model to make its heritage accessible, interoperable, and reusable, where permitted, following the FAIR (Findable, Accessible, Interoperable, Reusable) principles (Wilkinson et al. 2016).

Finally, it is appropriate to offer some concluding remarks regarding the applicability of this model for grouping and linking documents that are distant but related. Migrated Archives (Lowry and Chaterera-Zambuko 2023) is a specific example of an archive whose documents are often separated from their original fonds. In such cases, IIIF may be used to group these documents even in the digital environment. In conclusion, IIIF can be a turning point in the archival landscape, allowing users to access all the fonds in their original organization and preserving the archival bond.

Digitization is not just a technical issue concerning which tool to use and how to set it up. As Valacchi noted recently (2024; 18), it is important to rationalize the entire process in an anthropological and ethical way. In this context, archive digitization includes social, economic, and cultural aspects that need to be considered. For this reason, the role of archivists is essential to guarantee that technologies will be applied with respect to archives and aimed at promoting and valorizing them.

# Declarations

**Conflict of interest** The authors have no financial or proprietary interests in any material discussed in this article.

The archival material used for the development of the presented case study is entirely available and accessible online by FONDAZIONE ARCHIVIO AUDIOVISIVO DEL MOVIMENTO OPERAIO E DEMOCRATICO at: http://patrimonio.aamod.it/aamod-web/.

# References

1. Aly AS, Andrew C (2023) Practical Guide to Emergency Digitazion of Paper-Based Archival Heritage. Accessed 06 November 2025 at: https://www.ica.org/app/uploads/2024/07/ICA-Digitization-Manual-English-updated-12-07-2024_compressed.pdf
2. Archivio Centrale dello Stato (nd) Convegno "Dal caos al cosmo: la nuova teca digitale dell'Archivio centrale dello Stato". Arch. Cent. dello Stato https://acs.cultura.gov.it/convegno-dal-caos-al-cosmo-la-nuova-teca-digitale-dellarchivio-centrale-dello-stato/. Accessed 29 Oct 2025a
3. Archivio Centrale dello Stato (nd) inventario ID 0008 - Leggi e decreti dello Stato - Raccolta ufficiale delle leggi e dei decreti - RSI, 1943–1945. MLOL https://tecadigitaleacs.cultura.gov.it/item/d27232cd-13f8-4d61-b110-df7417a2ffac. Accessed 12 Apr 2025b
4. Archivio Storico della Gazzetta di Parma (nd) MLOL - Archivio storico della Gazzetta di Parma - il portale delle collezioni digitalizzate. Arch. Stor. della Gazz. di Parma https://archiviogazzettadiparma.medialibrary.it/home/index.aspx. Accessed 7 Apr 2025
5. Archivio Storico dell'Università di Bologna (nd) Patrimonio documentario - Archivio Storico. Arc. Sto. dell'Unive. di Bologna https://archiviostorico.unibo.it/it/patrimonio-documentario/patrimonio-documentario. Accessed 7 Apr 2025
6. Azim NAM, Yatin SFM, Jensonray RCA, Ayub@Mansor S (2018) Digitization of Records and Archives: Issues and Concerns. Int J Acad Res Bus Soc Sci 8: 170–178. https://doi.org/10.6007/IJARBSS/v8-i9/4582
7. Biblissima (nd) IIIF Collections of Manuscripts and Rare Books. Biblissima https://iiif.biblissima.fr/collections/about. Accessed 12 Apr 2025
8. Bird S, Klein E, Loper E (2009) Natural language processing with Python. O'Reilly, Cambridge

9. British Library (2025) Welcome to the Endangered Archives Programme. Endanger. Arch. Programme https://eap.bl.uk/welcome-endangered-archives-programme. Accessed 30 Oct 2025
10. British Library (nd) Native authority Mkumbira [1963–1964]. Endanger. Arch. Programme https://eap.bl.uk/archive-file/EAP427-2-9. Accessed 7 Apr 2025
11. Castellucci P (2017) Carte del nuovo mondo: banche dati e open access. Il mulino, Bologna
12. De Vincentis S, Critelli M (2025) Il “Moving Panorama” in IIIF. Usi di standard e di tecniche VR360 per un approccio alla museografia digitale. Um Digit 9:261–287. https://doi.org/10.6092/ISSN.2532-8816/21253
13. Direzione Generale Archivi (nd) Portale Antenati: Gli Archivi per la ricerca anagrafica. DGA https://archivi.cultura.gov.it/strumenti-di-ricerca-online/i-portali-tematici/portale-antenati. Accessed 12 Apr 2025
14. Freire N, Robson G, Howard JB, et al (2017) Metadata Aggregation: Assessing the Application of IIIF and Sitemaps Within Cultural Heritage. In: Kamps J, Tsakonas G, Manolopoulos Y, et al. (eds) Research and Advanced Technology for Digital Libraries. Springer International Publishing, Cham, pp 220–232
15. Geourjon A (nd) Mirador-annotation-editor-video. GitHub https://github.com/TETRAS-IIIF/mirador-annotation-editor-video. Accessed 06 Nov 2025.
16. Google (nd) Open Images Dataset V7 and Extensions. Google APIs https://storage.googleapis.com/openimages/web/index.html. Accessed 06 Nov 2025.
17. Grootendorst M (2022) BERTopic: Neural topic modeling with a class-based TF-IDF procedure. 10.48550/arXiv.2203.05794
18. Ines Montani, Matthew Honnibal, Adriane Boyd, et al (2023) explosion/spaCy: v3.7.2: Fixes for APIs and requirements. Accessed 06 November 2025 at: https://zenodo.org/doi/10.5281/zenodo.1212303
19. International Image Interoperability Framework (nd) Presentation API 3.0. IIIF https://iiif.io/api/presentation/3.0/. Accessed 12 Dec 2024a
20. International Image Interoperability Framework (nd) Presentation API 3.0 Change Log. IIIF https://iiif.io/api/presentation/3.0/change-log/. Accessed 12 Dec 2024b
21. International Image Interoperability Framework (nd) Change Discovery API 1.0. IIIF https://iiif.io/api/discovery/1.0/. Accessed 12 Apr 2025c
22. ISO/TR 13028 (2010) Information and documentation - Implementation guidelines for digitization of records
23. Istituto Centrale per la Digitalizzazione del Patrimonio Culturale – Digital Library (2022) Linee guida per la digitalizzazione del patrimonio culturale. Accessed 06 November 2025 at: https://digitallibrary.cultura.gov.it/wp-content/uploads/2023/06/PND-allegato1.pdf
24. Lodolini E (1995) Respect des fonds et principe de provenance: histoire, théories, pratiques. Gaz Arch 168:201–212. https://doi.org/10.3406/gazar.1995.4283

25. Lowry J, Chaterera-Zambuko F (2023) Lost Unities. In: The materiality of the migrated archives, 1st edn. Routledge, London
26. Manoni P (2023) AI4MSS: un esperimento di intelligenza artificiale alla Biblioteca Apostolica Vaticana. In: Guardando oltre i confini Partire dalla tradizione per costruire il futuro delle biblioteche Studi e testimonianze per i 70 anni di Mauro Guerrini. Associazione Italiana Biblioteche, Roma, pp 231–244
27. Matienzo M, Villa C, Keck J, et al (2024) ProjectMirador. GitHub https://github.com/ProjectMirador. Accessed 12 Dec 2024
28. Media Library Online Storie (nd) MLOL. MLOL https://educatt.medialibrary.it/pagine/pagina.aspx?id=680. Accessed 27 Jan 2025
29. Ministero della Cultura (2021) Online da oggi Alphabetica, il nuovo portale per esplorare il patrimonio culturale delle biblioteche italiane. Minist. Della Cult. https://cultura.gov.it/comunicato/21880. Accessed 5 May 2025
30. Ministero della Cultura (2023), Istituto per la digitalizzazione del patrimonio culturale - Digital Library Piano Nazionale di Digitalizzazione del Patrimonio Culturale 2022–2023. Versione 1.1. Accessed 06 November 2025 at: https://digitallibrary.cultura.gov.it/wp-content/uploads/2023/10/PND_V1_1_2023-1.pdf
31. Nockels J (2025) Making the past readable: a study of the impact of handwritten text recognition (HTR) on libraries and their users. https://era.ed.ac.uk/handle/1842/43452
32. Redmon J, Divvala S, Girshick R, Farhadi A (2016) You Only Look Once: Unified, Real-Time Object Detection. In: 2016 IEEE Conference on Computer Vision and Pattern Recognition (CVPR). IEEE, Las Vegas, NV, USA, pp 779–788
33. Robinson G, Companjen B, McCan P, et al (2020) Simple Annotation Server. GitHub https://github.com/glenrobson/SimpleAnnotationServer. Accessed 06 Nov 2025.
34. Romein CA, Rabus A, Leifert G, Ströbel PB (2025) Assessing advanced handwritten text recognition engines for digitizing historical documents. Int J Digit Humanit 7:115–134. https://doi.org/10.1007/s42803-025-00100-0
35. Şentürk B (2014) Effective Digitization in Archives. J Balk Libr Union 2:11. https://doi.org/10.16918/bluj.78275
36. Snydman S, Sanderson R, Cramer T (2015) The International Image Interoperability Framework (IIIF): A community & technology approach for web-based images. Arch Conf 12:16–21. https://doi.org/10.2352/issn.2168-3204.2015.12.1.art00005
37. Spina S (2023) Handwritten Text Recognition as a digital perspective of Archival Science. AIDAinformazioni 41:115–132. https://doi.org/10.57574/596529286
38. Terras M (2022) The role of the library when computers can read: Critically adopting Handwritten Text Recognition (HTR) technologies to support research. In: The rise of AI: implications and applications of artificial intelligence in academic libraries. Association of College and Research Libraries, Chicago, Illinois, pp 137–148

39. UniversalViewer (nd) UV. https://universalviewer.io/. Accessed 06 Nov 2025.
40. Valacchi F (2024) L’archivio aumentato: Tempi e modi di una digitalizzazione critica. edigita
41. W3C (nd) Web architecture: Metadata. W3C https://www.w3.org/DesignIssues/Metadata.html. Accessed 8 Apr 2025
42. Wilkinson MD, Dumontier M, Aalbersberg IjJ, et al (2016) The FAIR Guiding Principles for scientific data management and stewardship. Sci Data 3:160018. https://doi.org/10.1038/sdata.2016.18
43. Zherdeva YuA, Cherkasova MV, Sumburova EI (2020) Social Aspects Of Digitalization: Archive And Memory In The Digital Era. pp 541–547

# Figures

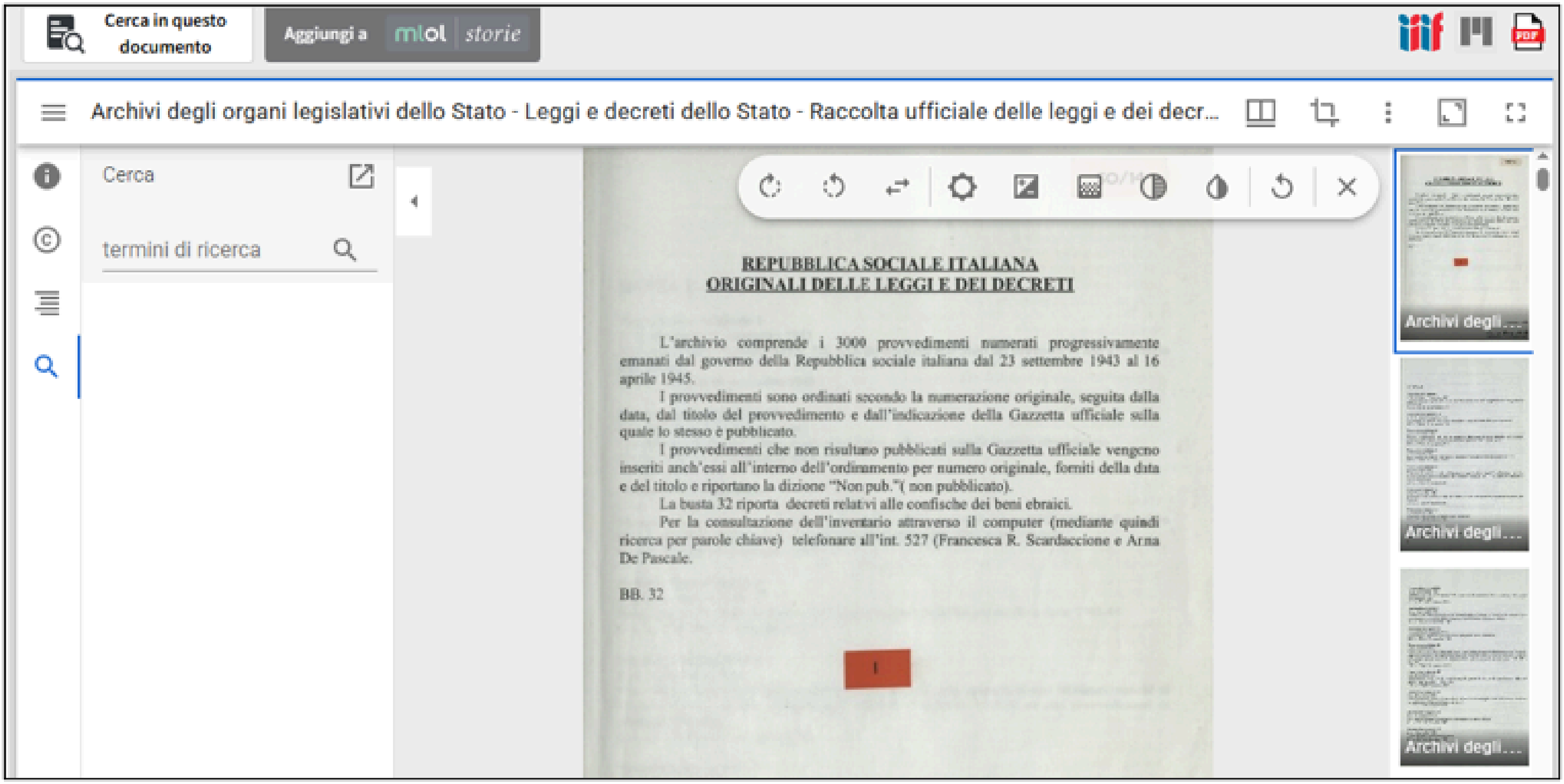

### Figure 1

Mirador 3 search terms screen for an ACS document (Archivio Centrale dello Stato).

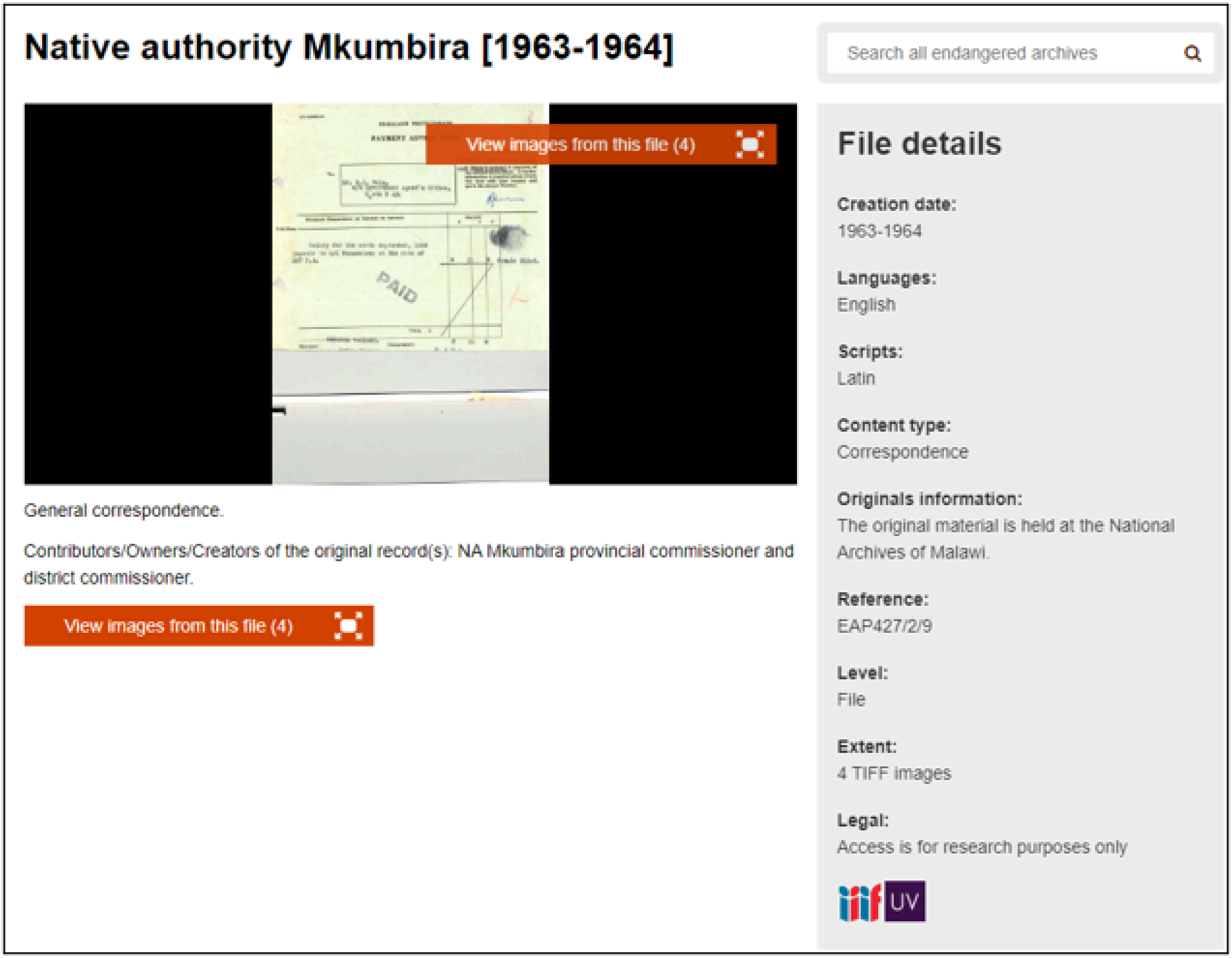

**Figure 2**

Document access screen (British Library)

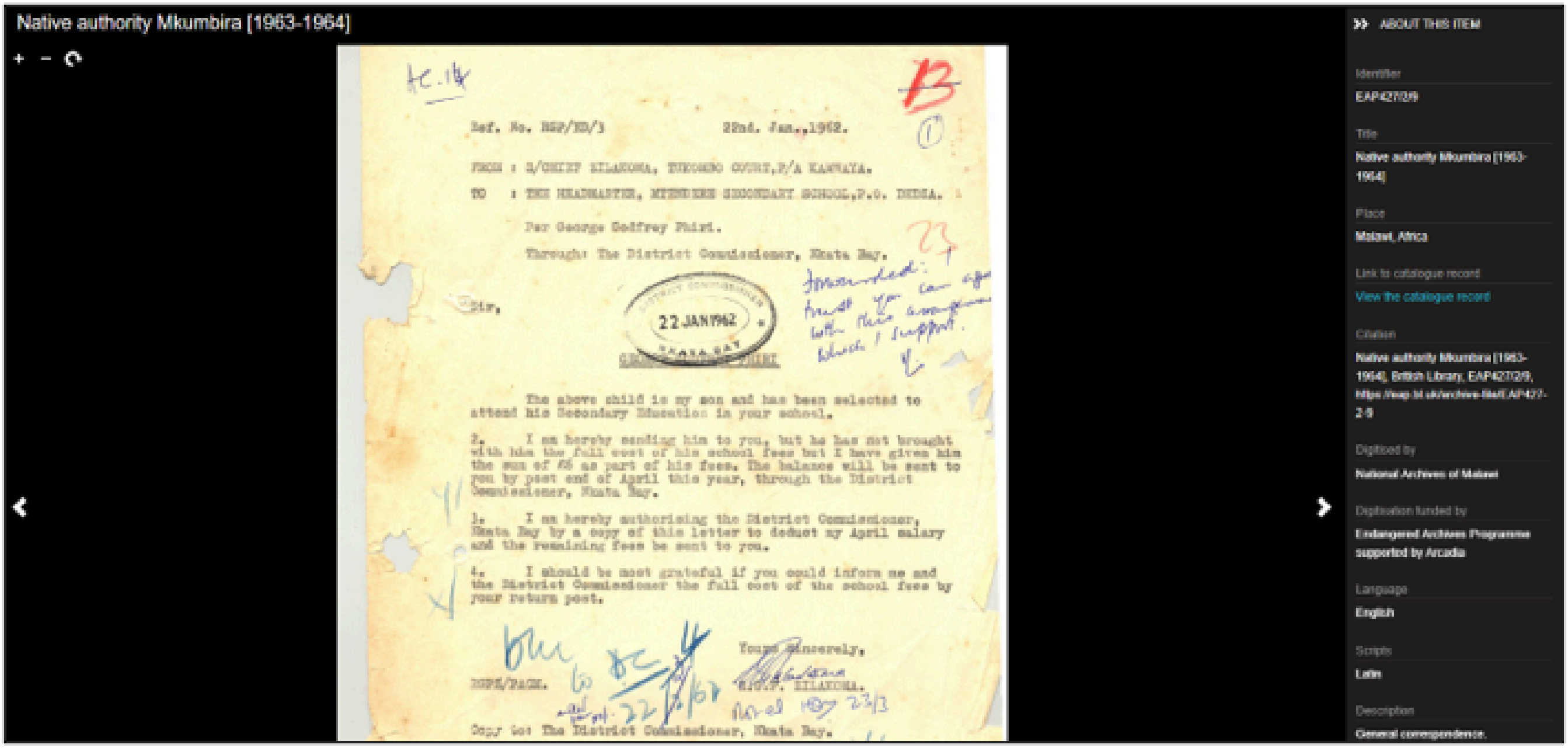

### Figure 3

Document viewing screen using Universal Viewer software directly accessible from the EAP website (British Library).

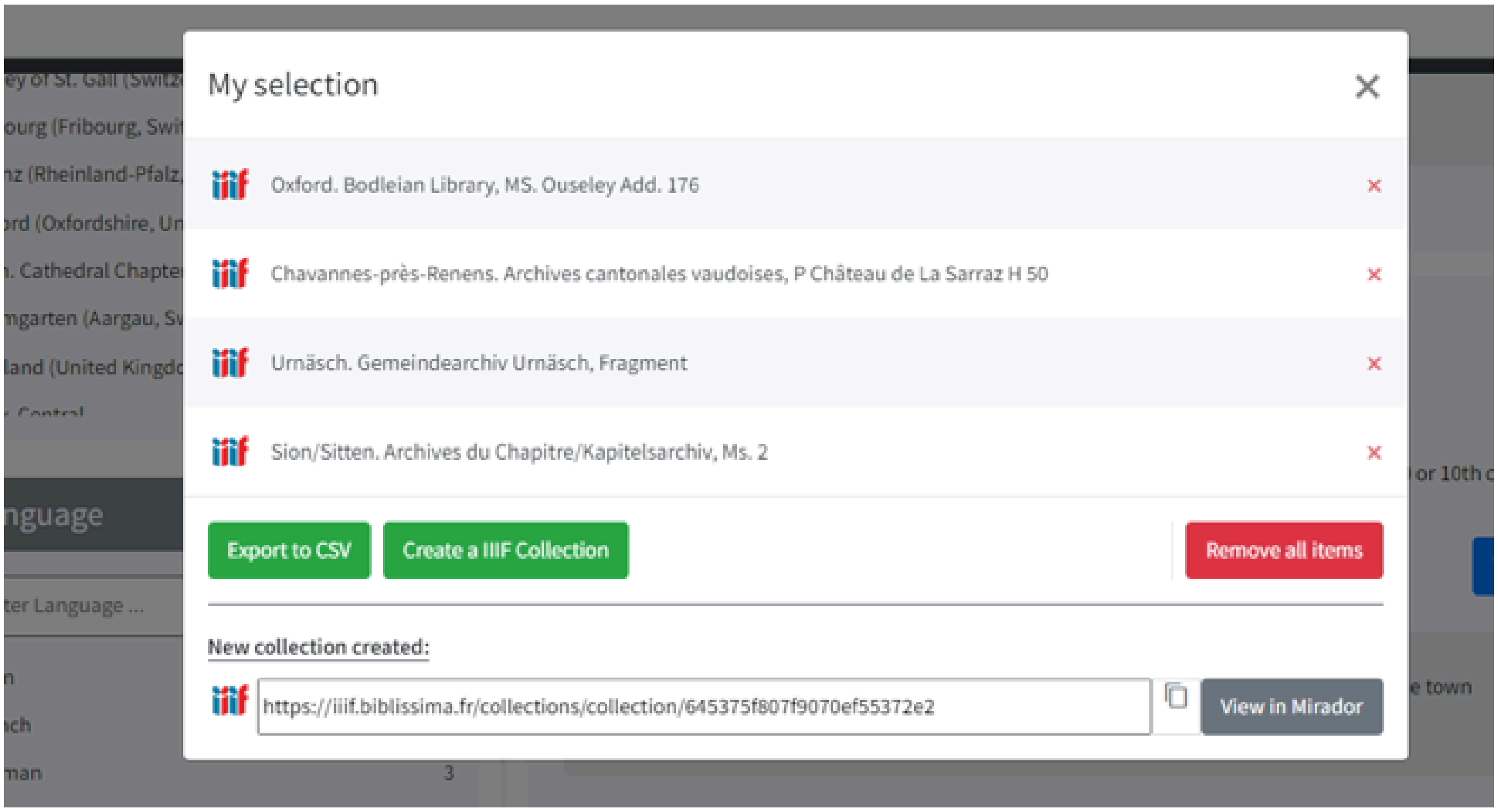

### Figure 4

Selection and creation of the IIIF collection (Biblissima).

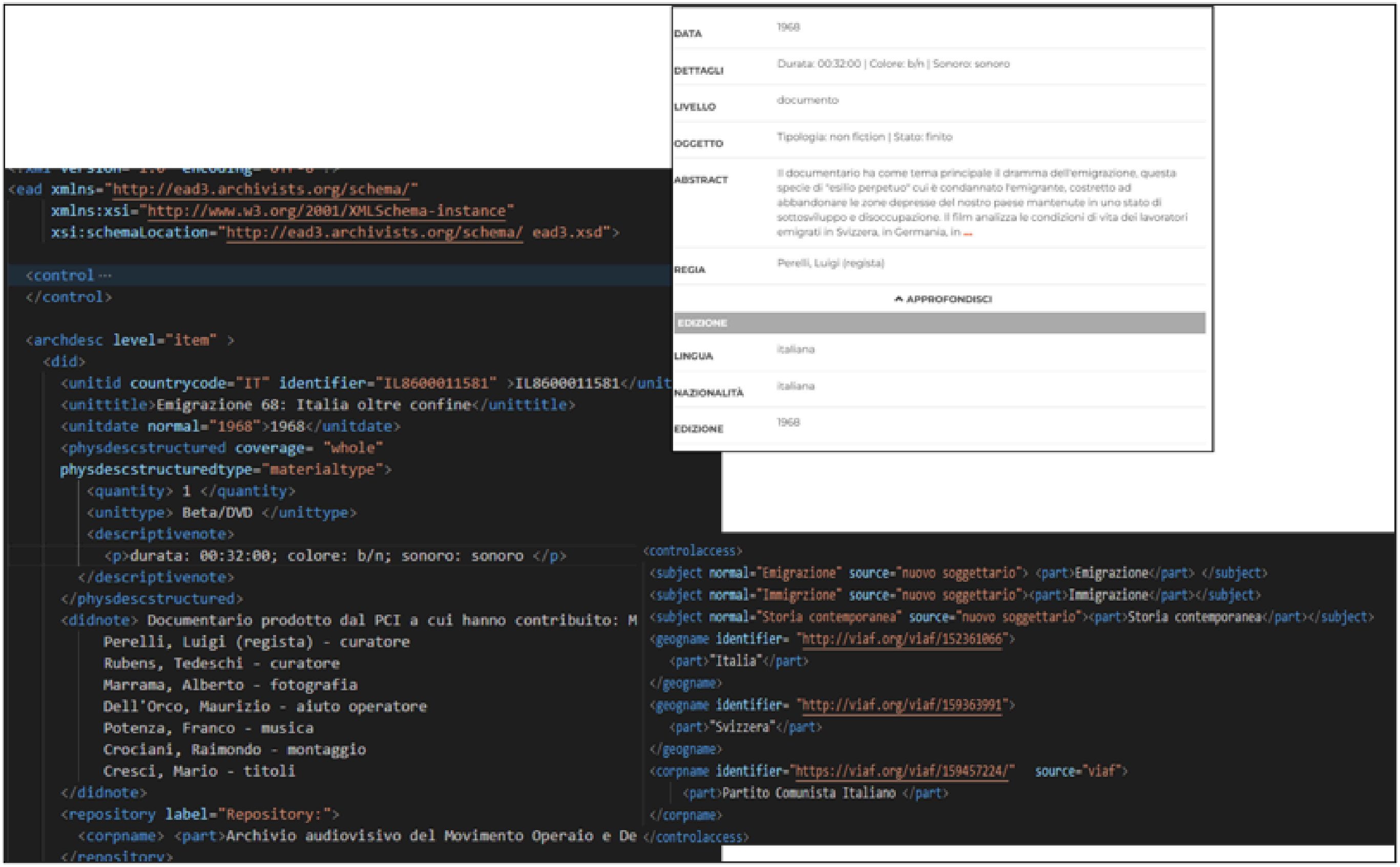

## Figure 5

Conversion of the description according to the EAD standard.

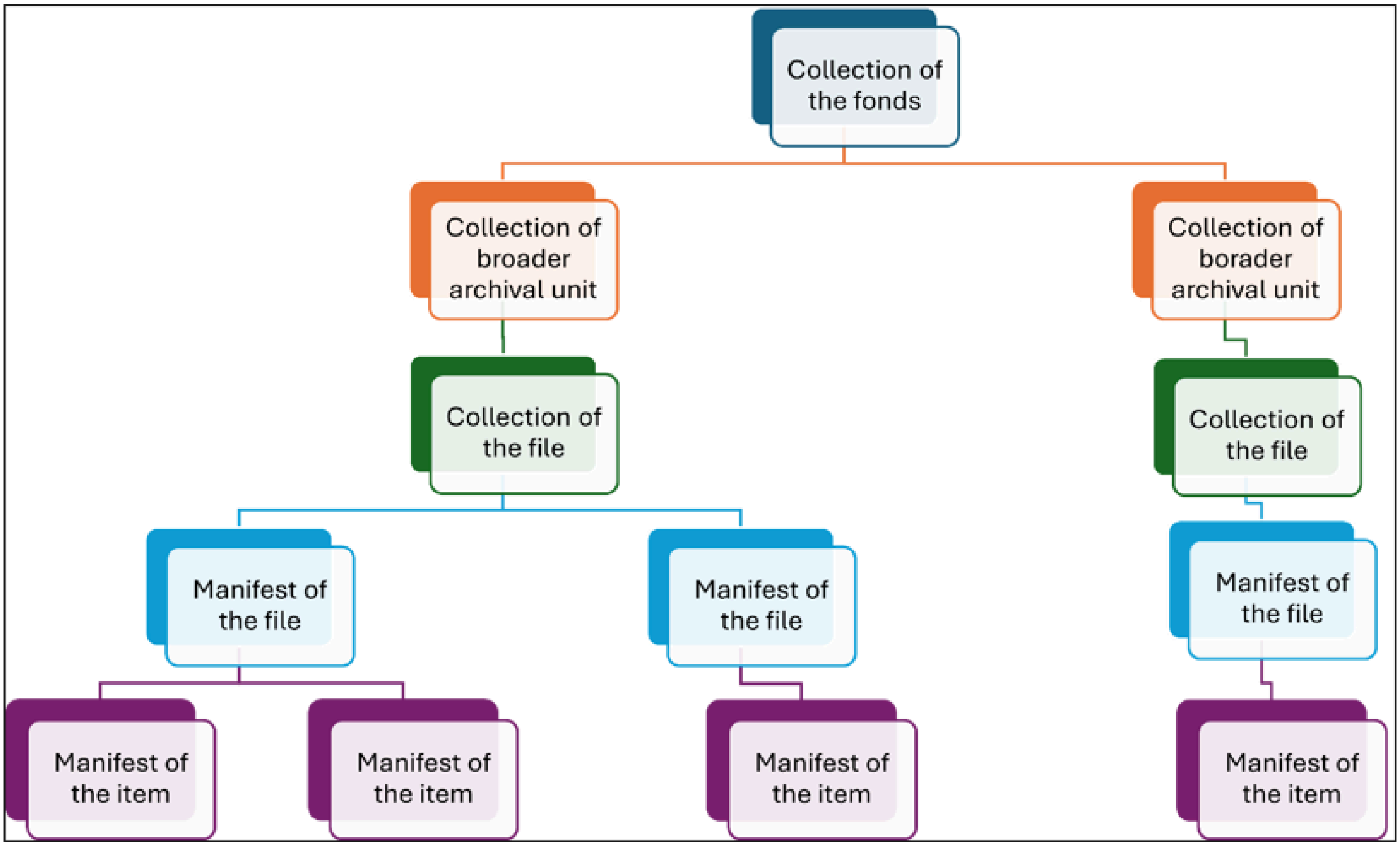

**Figure 6**

IIIF elements configuration to represent the archival hierarchy.

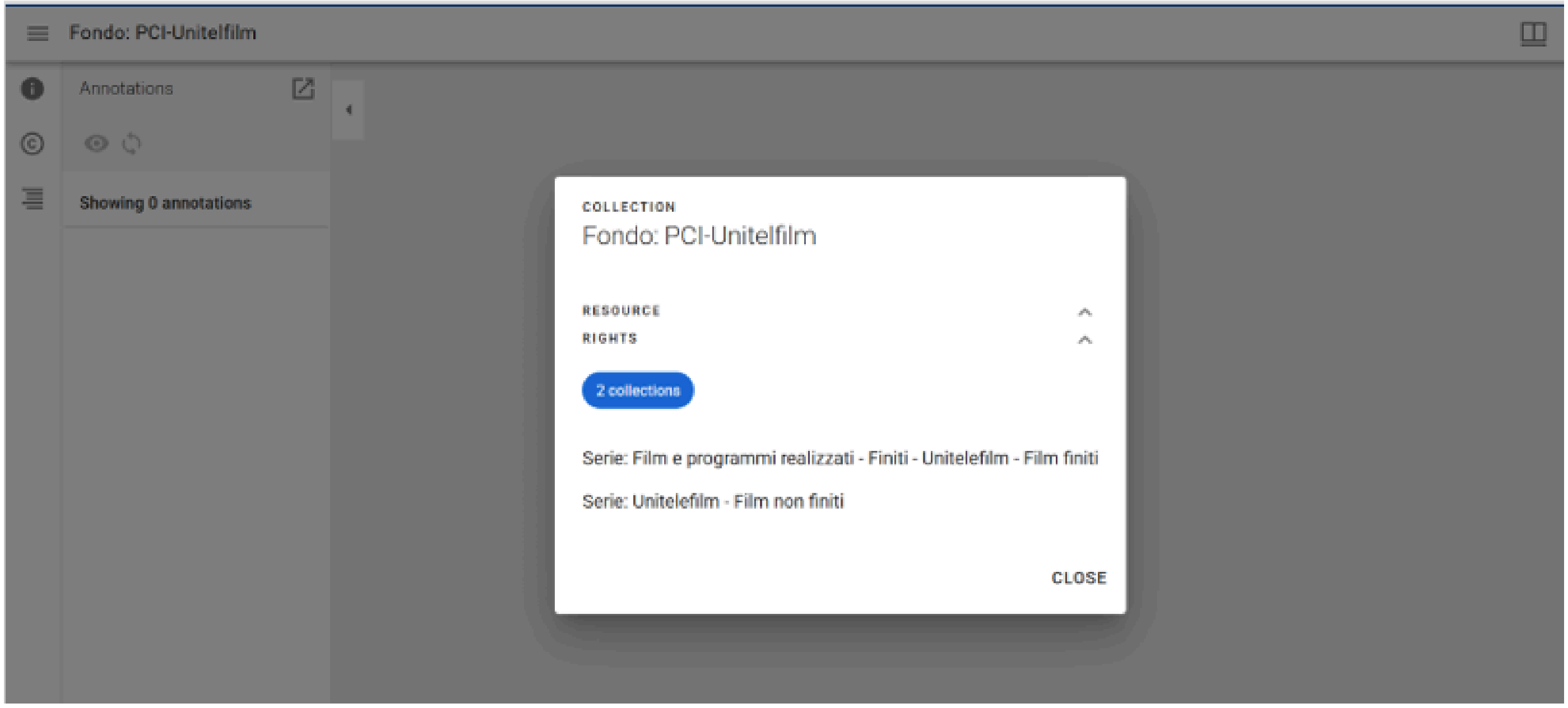

**Figure 7**

Home screen in Mirador 3 of the configured IIIF model.

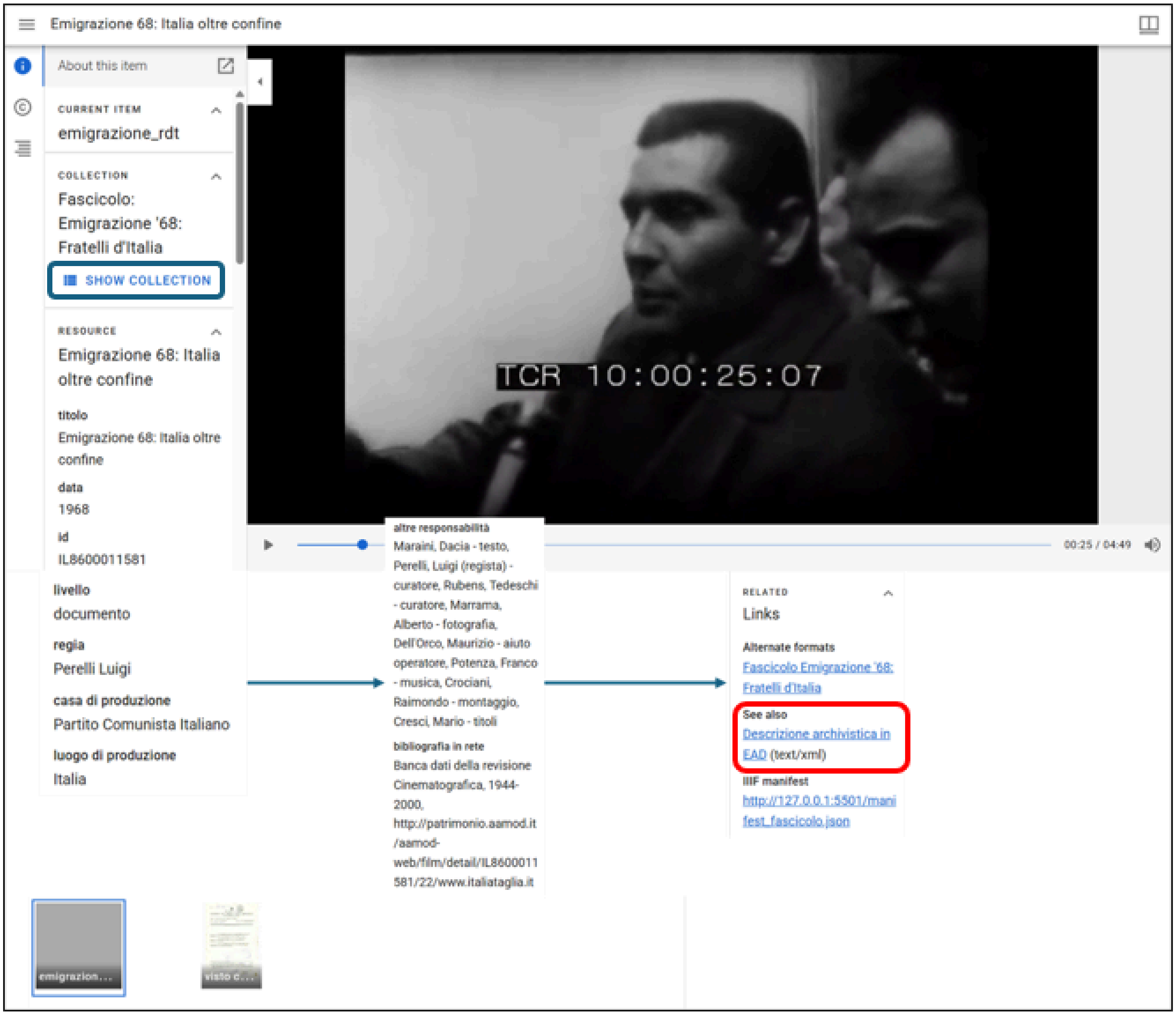

## Figure 8

Representation at the item level with the resource's metadata.

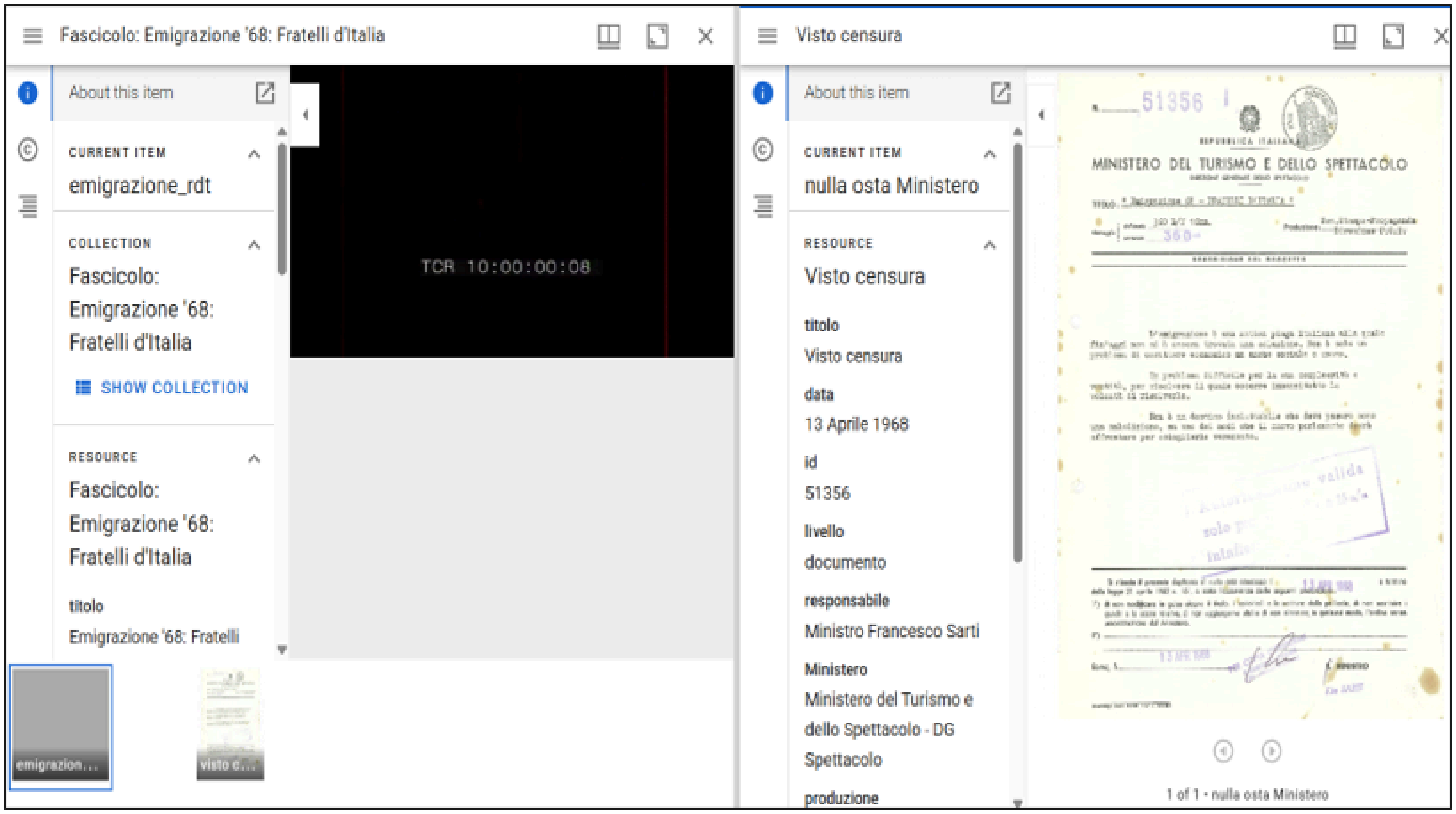

## Figure 9

Multiple objects view in Mirador 3.